\documentstyle[boxedepsf]{elsart}
\HideDisplacementBoxes
\begin{document}

\begin{flushright}
hep-ph/9611213 \\
IHEP 96--79 \\
September 1996
\end{flushright}

\vskip2cm

\begin{frontmatter}
\title{Extraction of parton distributions and $\alpha_s$ from DIS data
within the Bayesian treatment of systematic errors}
\author{Sergey Alekhin}
\address{{\tt alekhin@mx.ihep.su}\\
          Inst. for High Energy Physics, Protvino, 142284, Russia}

\begin{abstract}
We have performed the NLO QCD global fit of BCDMS, NMC, H1 and ZEUS data
with full account of point-to-point correlations
using the Bayesian approach to the treatment of systematic errors.
Parton distributions in the proton associated with experimental uncertainties,
including both statistical and systematic ones were obtained.
The gluon distribution in the wide region of $x$ was determined and
it turned out to be softer than in the global analysis using prompt
photon data. We also obtained the
robust estimate of $\alpha_s(M_Z)=0.1146\pm0.0036~(75\%~C.L.)$
based on Chebyshev's inequality, which
is compatible with the earlier determination
of $\alpha_s$ from DIS data, but with less dependence on high twist effects.
\end{abstract}
\end{frontmatter}
\clearpage

\section{Introduction}
Recently it has been argued \cite{SOPER}
that parton distributions functions (PDFs)
obtained from the global data analysis
(e.g. \cite{MRS,CTEQ}) have the principal shortcomings arising from
the absence of experimental
errors associated with the parameters of these distributions.
Indeed, the only and often used way
to evaluate the spread of predictions
given by these PDFs is to compare
results of calculations with the various parametrizations input.
It is evident that if different authors use the same theoretical
model and similar data sets this procedure cannot
account for real uncertainties occuring due to statistical
and systematic fluctuations of data used to extract PDFs.
These uncertainties can be
evaluated using the propagation of these fluctuations into
the dispersion of PDFs parameters or PDFs themselves.
The conclusive treatment of systematic errors, which are usually dominating,
is often limited since they are presented in the publications
as the combinations from separate sources. For the recent
deep inelastic scattering (DIS) data from HERA as well as older ones
from SPS full error matrix are fortunately available. Deep inelastic
scattering of charged leptons
remains the cleanest source of information on PDFs among
the other relevant processes and the careful analysis of these data
including propagation of systematics can be valuable for exploring
the nucleon structure.
The handling with statistical fluctuations
is well understood on the basis of probability theory, meanwhile
the elaborating of systematic ones
is the subject of various approaches. 

In one of them, based on the classical treatment of probability, one considers 
the systematic shifts as additional unknown methodical
parameters arising due to 
a poor knowledge of experimental apparatus. Within this approach one 
usually tries to determine these parameters using some statistical 
estimator, say  $\chi^2$ minimization,  
to fit the data with these parameters left free. 
The obtained values are further 
considered as a reasonable approximation to the true values and data are 
corrected to account for these systematic shifts. As to systematic errors
of theoretical model parameters, they are evaluated inverting full
error matrix, including both physical and methodical parameter derivatives.
In most cases, the only kinds of the systematic errors 
which can be determined in the pure 
classical approach are the systematics connected to the 
general normalization of the data.
Other methodical parameters are strongly correlated with each other
and with physical parameters which leads to their huge errors and 
unreasonable central values. This situation can be readily 
explained qualitatively: as far as one turned out to be unable to 
determine the parameters of the apparatus using the special tests and 
measurements it is doubtful that one can do it using some cross section 
measurements indirectly related to the resolving of the 
methodical ambiguities. 

Another, much more productive approach, is based on the Bayesian treatment
of systematic uncertainties. In this approach they are considered 
as random variables with the postulated/evaluated probability 
distribution function and systematic errors are evaluated within 
general statistical procedures alongside with the statistical errors. 
For the analysis of the modern DIS data as a rule having a number of
noticeable systematic errors this approach is the unique possibility 
to account for the point-to-point correlation of data.
This is the Bayesian approach that we use in our paper to obtain
the complete propagation of 
systematic uncertainties of DIS data into the uncertainties of 
the resulting PDFs.

\section{Theoretical and experimental input}
\subsection{Data used in the fit}
 As a subject of our analysis we use the data 
for deep inelastic muon/electron
hydrogen/deuterium scattering \cite{BCDMS,NMC,ZEUS,H1}
cut to reduce the effects of high twists in the following way
\begin{displaymath}
W~>~4~GeV,~~~~~Q^2~>~9~GeV^2,
\end{displaymath}
where $W$ and $Q^2$ are common DIS variables. 
The number of data points for each experiment after the cut
is presented in Table 1.
\begin{table}[h] 
\caption{The number of data points (NDP) and
$\chi^2$/NDP for the analysed data sets.} 
\begin{center}
\begin{tabular}{|c|c|c|c|c|c|c|c|}   \hline
Experiment&BCDMS&NMC&H1&ZEUS&total \\ \hline
NDP&558&190&147&166&1061    \\ \hline
$\chi^2$/NDP&0.97&1.43&0.91&2.00&1.20          \\ \hline
\end{tabular}
\end{center}
\end{table}
For data of ZEUS collaboration asymmetric systematic errors were averaged.
As to BCDMS data we suppose the total correlation of systematic errors
for proton and deuterium cross sections. 
\subsection{Probability model of the data}
If the experimental data with {K} sources of 
multiplicative systematics are explicitly described by a theoretical model
they can be presented in the Bayesian approach as
\begin{displaymath}
y_i=(f_i+\mu_i \sigma_i)\cdot(1 + \sum_{k=1}^{K}\lambda_k \eta_i^k),
\end{displaymath}
where $f_i=f_i(\theta^0)$ is the value predicted by the
theoretical model with parameter $\theta^0$,
$\mu_i$ and $\lambda_k$ are independent
random variables, $\sigma_i$ and $\eta_i^k$ -- statistic and
systematic errors from the $k$-th source for $i$-th measurement,
$i=1 \cdots N$, $k=1 \cdots K$, $N$ is the total number of points
in the data set. 
If the data come from the data sample with a large number of events
in every bin, $\mu$ are normally distributed,
as to $\lambda$, the only assumption we are making is that they have
zero average and unity dispersions. 
Within this ansatz 
individual measurements are correlated and their correlation matrix 
$C_{ij}$ is given by 
\begin{displaymath}
C_{ij} = \sum_{k=1}^K f_i \eta_i^k f_j \eta_j^k+\delta_{ij}\sigma_i^2
\end{displaymath}
where $\delta_{ij}$ is the Kronecker symbol.
To obtain the estimator of the parameter $\theta^0$ 
we minimize the quadratic form
\begin{equation}
\chi^2(\theta)=\sum_{i,j=1}^{N} [f_i(\theta)-y_i] E_{ij}  [f_j(\theta)-y_j],
\label{CHI2}
\end{equation}
where $E_{ij}$ is inverted correlation matrix.
We should note that through this paper
we treat the normalization errors within this formalism as well 
as other systematics are regarded as multiplicative, which is 
almost always the case for counting experiments.
The minimization was made with the help of MINUIT package \cite {MINUIT}
supplied with the modules improving the numerical stability
of calculations \cite{STAB}.

If $\lambda_k$ are normally distributed and $\eta_i^k\ll 1$,
$\lbrace y_i\rbrace$ set obeys the multidimensional 
Gaussian distribution with correlations
and $\hat\theta$ has the minimal possible dispersion. 
The systematic errors calculated as the propagation of 
uncertainties in apparatus parameters or Monte-Carlo corrections
are well believed to be Gaussian distributed.   
At the same time we have shown \cite{DISP} that even this is not the case
this estimator has 
reduced dispersion comparing with the simplest $\chi^2$ without 
account of correlations. One should underline that 
as far as we use the correct covariance matrix built using predicted 
averages for the measurements our estimator would be
asymptotically unbiased 
and hence does not suffer from the bias discussed in \cite{AGO}.
\subsection{QCD input}
Physical model for describing the considered data is based on
the parton model with pQCD evolution 
of the light quarks and gluon distributions in the proton 
defined at initial value of $Q_0^2=9~GeV^2$.
These distributions were evolved using DGLAP equations \cite{AP}
in the NLO within $\overline{MS}$ factorization scheme \cite{MS}. 
As to the contributions of $c$-quark and $b$-quark they were calculated
using the LO formula from \cite{GLUCK} setting $m_c=1.5~GeV$, $m_b=4.5~GeV$
and the renormalization/factorization scale equal to 
$\sqrt{Q^2+4m^2_{c,b}}$.
Our QCD evolution program was tested as suggested in \cite{NLOCODES}
and demonstrated numerical precision of $O(0.1\%)$
in the kinematic region covered by the analysed data.
Adjusting the functional form of PDFs we've started from rather general 
and widely used expressions
\begin{equation}
xq_i(x,Q_0)=A_ix^{a_i}(1-x)^{b_i}(1+\gamma^i_1\sqrt{x}+\gamma^i_2 x),
\label{GENPDF}
\end{equation}
and then reduced the number of free parameters keeping the 
quality of data description.
The resulting functional form of PDFs at $Q_0$ looks like
\begin{displaymath}
xd_V(x,Q_0)=\frac{1}{N^{V}_d}x^{a_d}(1-x)^{b_d},~
xd_S(x,Q_0)=\frac{A_S}{N_S}x^{a_{sd}}(1-x)^{b_{sd}},
\end{displaymath}
\begin{displaymath}
xu_{V}(x,Q_0)=\frac{2}{N^{V}_u}x^{a_u}(1-x)^{b_u}(1+\gamma_2^{u}x),~
xu_S(x,Q_0)=\frac{A_S}{N_S}\eta_ux^{a_{su}}(1-x)^{b_{su}},
\end{displaymath}
\begin{displaymath}
xG(x,Q_0)=A_Gx^{a_G}(1-x)^{b_G},~~~~~~
xs_S(x,Q_0)=\frac{A_S}{N_S}\eta_sx^{a_{ss}}(1-x)^{b_{ss}}.
\end{displaymath}
We did not consider
$N^{V}_u, N^{V}_d$ and $A_G$ as free parameters, they were 
calculated from other parameters using partons' number/momentum conservation.
As to $N_S$ it is defined by the relation
\begin{displaymath}
2\int_0^1x\bigl[u_s(x,Q_0)+d_s(x,Q_0)+s_s(x,Q_0)\bigr]dx=A_S.
\end{displaymath}
Forecasting the final results we note that after trial fits
it has been found that $\eta_u$ is well compatible with unity
and it is fixed at this value.
We fixed $\eta_s=0.5$,
which is compatible  with recent CCFR findings \cite{CCFR}
and also adopted $a_{su}=a_{sd}=a_{ss}$,
$b_{ss}=(b_{su}+b_{sd})/2$
since our data do not allow for a separate determination 
of these parameters.

We calculate strong coupling constant $\alpha_s(Q)$
from the fitted parameter $\alpha_s(M_Z)$  
by numerical solving of the NLO renormalization equation 
\begin{displaymath}
\frac{1}{\alpha_s(Q)}-\frac{1}{\alpha_s(M_Z)}=
\frac{\beta_0}{2\pi}\ln\biggl(\frac{Q}{M_Z}\biggr)+
\beta\ln\biggl[\frac{\beta+1/\alpha_s(Q)}{\beta+1/\alpha_s(M_Z)}\biggr],
\end{displaymath}
where
\begin{displaymath}
\beta_0=11-\frac{2}{3}n_f,~~~~\beta=\frac{2\pi\beta_0}{51-\frac{19}{3}n_f}.
\end{displaymath}
This approach prevents one from the uncertainties
occuring for the approximate solutions based on the expansion 
in the inverse powers of $\ln(Q)$,
which are $\sim0.001$ at the scale of evolution 
from $M_Z$ to $O(GeV)$ (cf. \cite{PDG}), i.e. is comparable with 
the standard deviation of $\alpha(M_Z)$.  
The number of the active fermions $n_f$ is changing from $4$ to $5$ 
due to b-quark threshold at the $Q=m_b$
keeping continuity of $\alpha_s(Q)$.

\subsection{Corrections to the basic formula and data}
\subsubsection{Target mass correction}
 In addition to the pure pQCD evolution we applied to the calculated 
value of $F_2$ the so-called target mass corrections \cite{TMC} using the 
relation
\begin{displaymath}
F_2^{TMC}(x,Q)=\frac{x^2}{\tau^{3/2}}\frac{F_2(\xi,Q)}{\xi^2}
+6\frac{M^2}{Q}\frac{x^3}{\tau^2}\int^{1}_{\xi}dz\frac{F_2(z,Q)}{z^2},
\end{displaymath}
where
\begin{displaymath}
\xi=\frac{2x}{1+\sqrt{\tau}},~~~~\tau=1+\frac{4M^2x^2}{Q^2}
\end{displaymath}
and $M$ is the nucleon mass. The contribution to this correction of the 
order of $M^4/Q^4$ presented in \cite{TMC}
turned out to be negligible for all considered data.
Target mass correction is most essential for the BCDMS data,
where it ranges from --1\% to +7\%, having the average module related 
to the statistical error as large as 0.16. We should note that 
our way of introducing this correction differs from the one
applied in \cite{LOWQ} and consisting of the substitution 
$F_2(x,Q)\rightarrow F_2(\xi,Q)$. Due to this difference in our 
case the correction exhibits crossover from negative to positive
values at $x\approx0.5$ instead of $x\approx0.4$ like in \cite{LOWQ}
and differs in the magnitude. 
For the NMC data this correction is significantly smaller 
( range -- [--1\%,0\%], relative average -- 0.05 ) and for ZEUS and H1
data is absolutely negligible.
\subsubsection{Reduction to the common $R=\sigma_L/\sigma_T$}
All the data on $F_2$ were 
reduced to the common value of $R=\sigma_L/\sigma_T$
comprised the NLO contribution from light quarks and gluon,
the LO contribution from $c$-quark and $b$-quark, 
and the target mass correction included
(see \cite{WHIMP} for the compilation of the relevant formula). 
The value of $R$
was calculated during the fit for every new set of the 
PDFs parameters (its final form is presented on Fig.1).
\begin{figure}[t]
\begin{center}
\ForceHeight{0.4\textheight}\BoxedEPSF{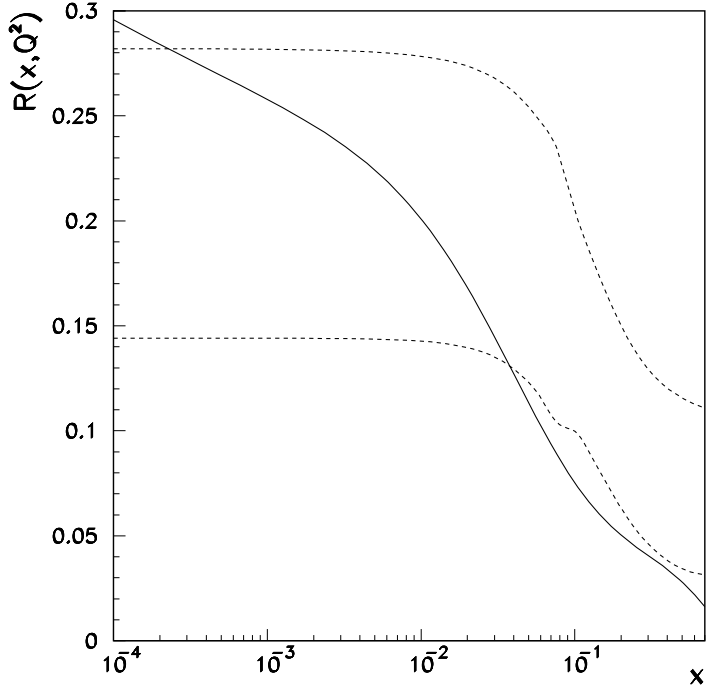 scaled 1000}
\caption{$R=\sigma_L/\sigma_T$ calculated using our resulting PDFs
(solid line) and the band of $R^{1990}_{SLAC}$ [22]
(dashed lines) at $Q^2=9~GeV^2$.}
\end{center}
\end{figure}
This reduction is most essential at the smallest $x$ accessible 
in an experiment, mainly, due 
the maximum sensitivity of the data to the value of $R$ in these regions.
The value of this correction is different for the considered data sets.
For the BCDMS data the value of 
this correction is in the range of [--3.5\%,0\%] (the average 
relative module -- 0.10). 
This collaboration calculated $R$ from pQCD predictions, but 
used the larger gluon distributions than in our final set.
The NMC data are renormalized by 0.10 statistical error in average 
(range -- [--1.5\%,2\%]).
For the ZEUS data, which exhibit the most sensitivity to the choice of $R$
due to the large span in lepton scattering variable $y$,
 this correction calculated with the final set 
of our PDFs ranges from --3\% to 0\% with the average relative module 
of 0.04 and as to the H1 data they are affected to the same extent.

Resuming we should note that this correction, being not very 
large in average, is significant for separate data points
on the edge of the experimental acceptance. 
Since at small $x$ the value of $R$ heavily depends on the 
$G(x,Q)$, our approach imposes the additional constraints
on its value.
The residual influence of different ansatzes for $R$ used 
in the calculation of radiative corrections in different experiments
is believed to be small.

\subsubsection{Fermi motion correction in deuterium}
Deuterium data were corrected for Fermi motion using
procedure \cite{WEST} with the Paris wave function for
deuterium \cite{PARIS}. This correction was also calculated iteratively
to obtain fully consistent set of PDFs. 
The value of $R=\sigma_L/\sigma_T$ for deuteron was adopted to be unchanged 
under this correction, we have proved that this adoption is of 
minor importance for the final results.
 For the calculation of the relevant
integrals we used program \cite{SOKOL}, which exhibited 
better numerical stability than standard procedures based on the simple
Gauss algorithm. This correction being maximum at large $x$
ranges from --2\% to +15\% for the BCDMS data and from --2\% to --1\%
for the NMC data, whereas its average relative module 
is about 0.6 for the both experiments.
\begin{table}[t] 
\caption{The fitted parameters of PDFs with the full experimental errors
including statistics and systematics.}
\begin{center}
\begin{tabular}{|c|c|c|}   \hline
Valence&$a_u$&$0.745\pm0.024$        \\\cline{2-3}
       &$b_u$&$3.823\pm0.070$        \\\cline{2-3}
       &$\gamma_2^u$&$0.56\pm0.28$    \\\cline{2-3}
       &$a_d$&$0.875\pm0.066$        \\\cline{2-3}
       &$b_d$&$5.32\pm0.22$        \\\hline 
Glue   &$a_G$&$-0.267\pm0.043$        \\\cline{2-3}
       &$b_G$&$8.2\pm1.5$        \\\hline 
\end{tabular}
\begin{tabular}{|c|c|c|}   \hline
Sea    &$A_S$&$0.159\pm0.036$        \\\cline{2-3}
       &$a_{sd}$&$-0.1885\pm0.0072$        \\\cline{2-3}
       &$b_{sd}$&$7.5\pm1.3$        \\\cline{2-3}
       &$\eta_{u}$&$1.0\pm0.12$        \\\cline{2-3}
       &$b_{su}$&$10.61\pm0.95$        \\\cline{2-3} 
       &$\eta_{s}$&$0.5\pm1.0$        \\\hline 
       &$\alpha_s(M_Z)$&$0.1146\pm0.0018$  \\\hline 
\end{tabular}
\end{center}
\end{table}
\section{Results}
\begin{figure}[t]
\begin{center}
\ForceHeight{0.6\textheight}\BoxedEPSF{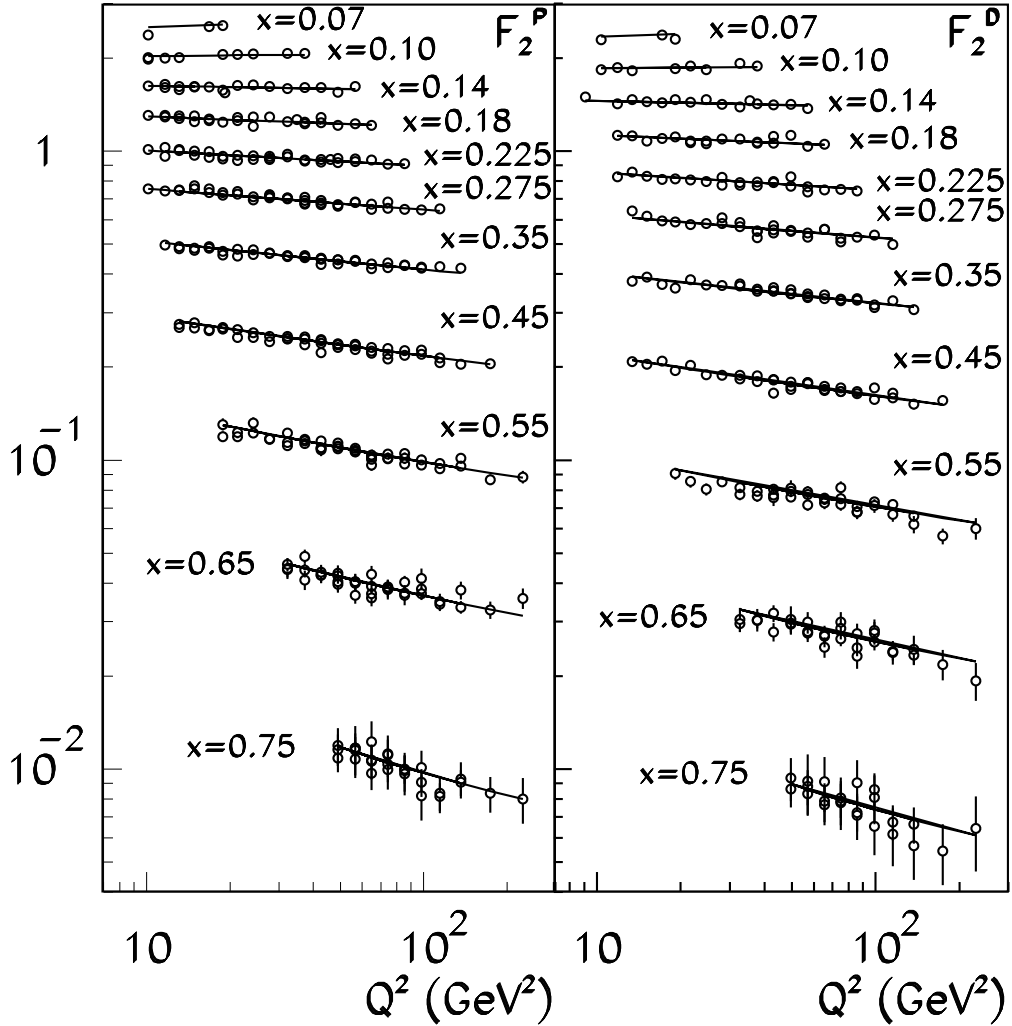 scaled 1000}
\caption{The description of BCDMS data with our PDFs. The data and curves
are scaled by factor $1.2^{11-i}$, where $i$ runs from 1 for the highest x bin 
to 11 for the lowest one.}
\end{center}
\end{figure}
The central values and the full experimental errors 
of the adjustable parameters obtained after the minimization
of (\ref{CHI2}) are presented in Table 2
(full correlation matrix of the fitted parameters is available 
by the request to the author).
To decrease the model dependence 
of our predictions, calculating the covariance matrix we 
released parameters $\eta_u$ and $\eta_s$, keeping their 
central values intact. 
The resulting $\chi^2$ values are presented in Table 1. 
On the average the model describes the data fairly well. One can heavily
ascribe rather large
$\chi^2$ obtained for the NMC and ZEUS data to the shortcoming of 
the theoretical model, as far as as the BCDMS and H1 data
having comparable statistics and lying in the nearby kinematic 
regions are described by this model perfectly.
The most probable explanation is that some systematic errors in these 
experiments are not Gaussian distributed. 
The average bias of the 
data against our model, calculated as
\begin{displaymath}
B=\Biggl\langle\frac{f-y}{\sqrt{\sigma^2+f \sum_{k=1}^{K}(\eta^k)^2}}
\Biggl\rangle
\end{displaymath}
turned out to be 0.10, i.e. is statistically insignificant.
The principal difference of our analysis from other global fits 
is that we do not renormalize data and as far the BCDMS data are usually 
shifted down, our resulting $F_2$ curves are slightly higher than others
at large $x$.
\begin{figure}[t]
\begin{center}
\ForceHeight{0.6\textheight}\BoxedEPSF{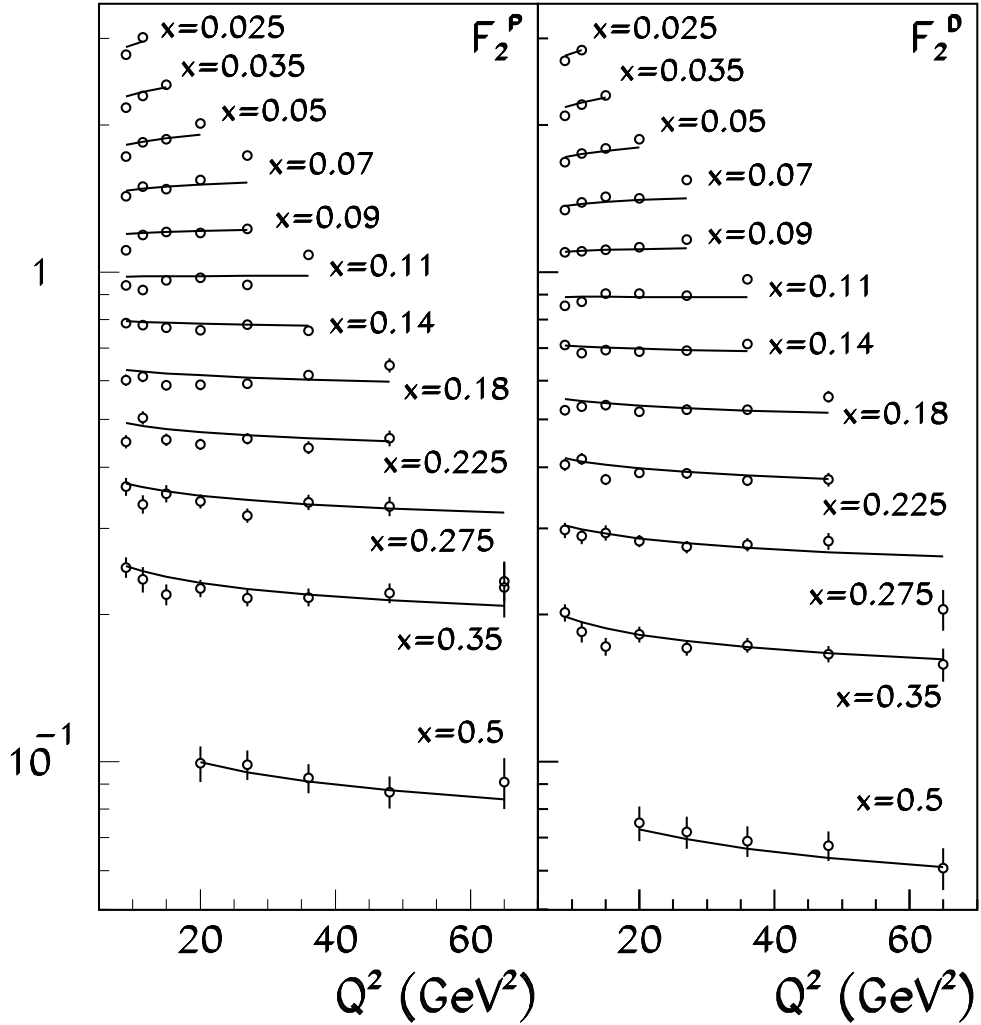 scaled 1000}
\caption{The same as in Fig.2 for the NMC data
($i$ runs from 1 to 12). For the presentation 
purposes we pictured combined energy data with convenient binning.}
\end{center}
\end{figure}
The data on $F_2$ reduced to the common value of $R$
together with our curves are presented on Figs.2--5, where 
the error bars correspond to the squared sum of statistics and systematics.
The selected set of PDFs is presented on Figs.6--9. 
The strange sea is not shown 
since from the analysed data we can obtain only a weak upper limit for 
this value. As we have mentioned 
above the distribution of our PDFs parameters defined mainly 
by the distribution of systematic uncertainties
 may differ from Gaussian
and then for the robust error bands estimate one should better use
Chebyshev's inequality.
The bands presented on these pictures 
correspond to two standard deviations, which corresponds to the 
75\% robust confidence level.
Although we do not use in our analysis prompt photon data,
which is often considered as an unique source of gluon distribution
at moderate $x$, through the kinematic region of $x=[0.0001,0.5]$
gluon distributions is determined rather precisely
and better than in the earlier analysis \cite{H1,NMCGLU}.
\begin{figure}[t]
\begin{center}
\ForceHeight{0.394\textheight}\BoxedEPSF{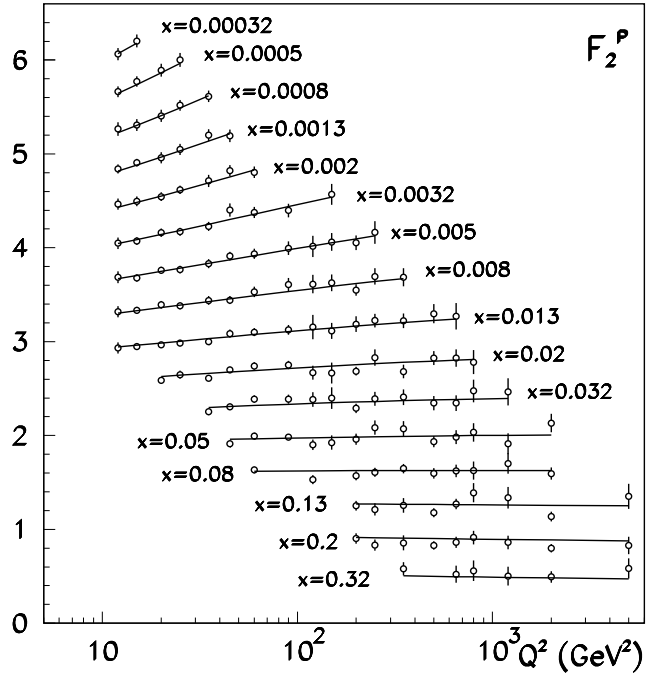 scaled 1000}
\caption{The description of H1 data with our PDFs. The data and curves
are shifted by $5.1-0.3i$, where $i$ runs from 1 for the highest x bin 
to 16 for the lowest one.}
\end{center}
\end{figure}
\begin{figure}
\begin{center}
\ForceHeight{0.394\textheight}\BoxedEPSF{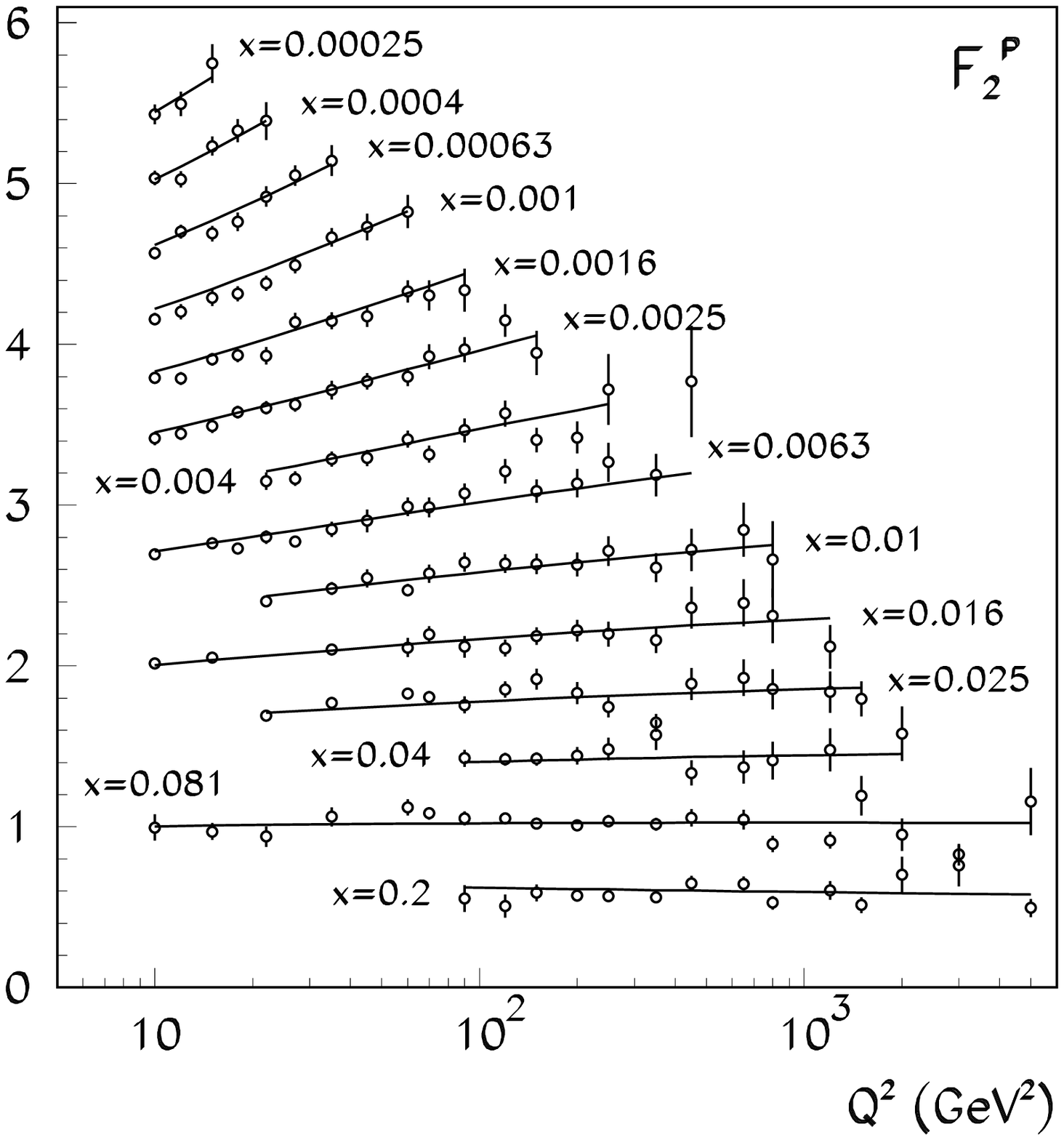 scaled 1000}
\caption{The description of ZEUS data with our PDFs. The data and curves
are shifted by $4.5-0.3i$, where $i$ runs from 1 for the highest x bin 
to 14 for the lowest one.}
\end{center}
\end{figure}
\begin{figure}
\begin{center}
\ForceHeight{0.4\textheight}\BoxedEPSF{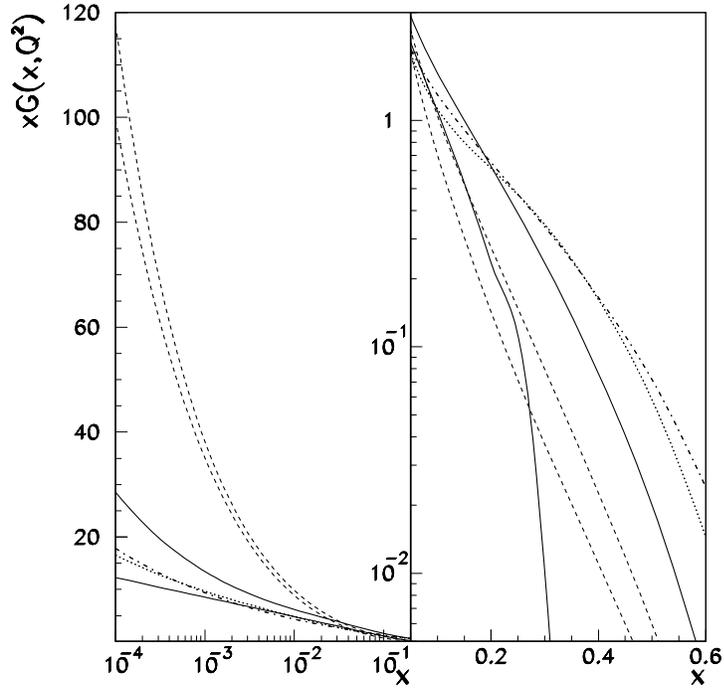 scaled 1000}
\caption{Gluon distribution obtained in our analysis.
Solid lines correspond to $Q^2=9~GeV^2$, dashed -- to $Q^2=10000~GeV^2$. 
Dotted line gives MRS(R1) and dashed-dotted -- CTEQ4M 
predictions at $Q^2=9~GeV^2$.}
\end{center}
\end{figure}
\begin{figure}
\begin{center}
\ForceHeight{0.44\textheight}\BoxedEPSF{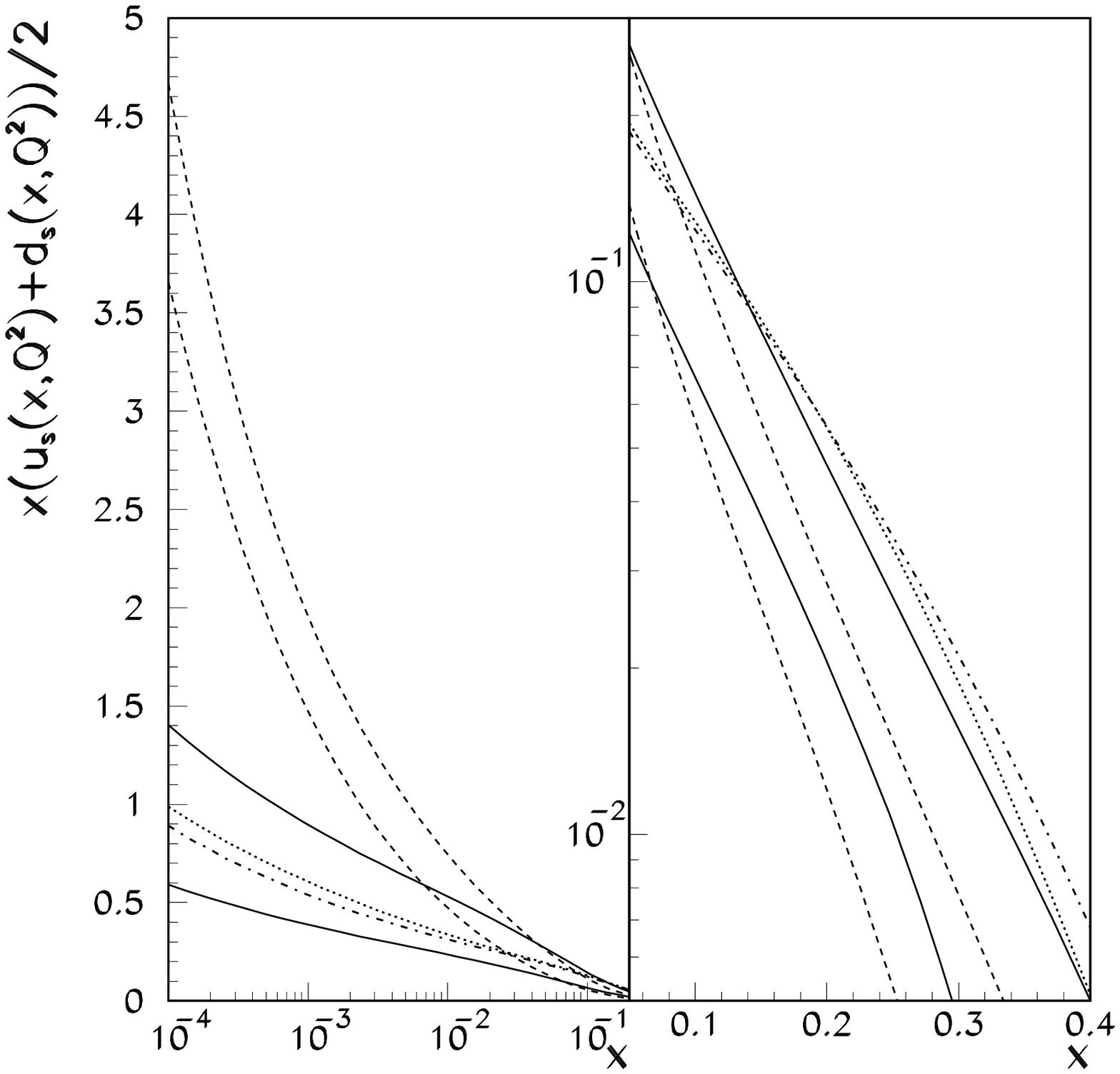 scaled 1000}
\caption{The same as in Fig.6 for the nonstrange sea.}
\end{center}
\end{figure}
\begin{figure}
\begin{center}
\ForceHeight{0.44\textheight}\BoxedEPSF{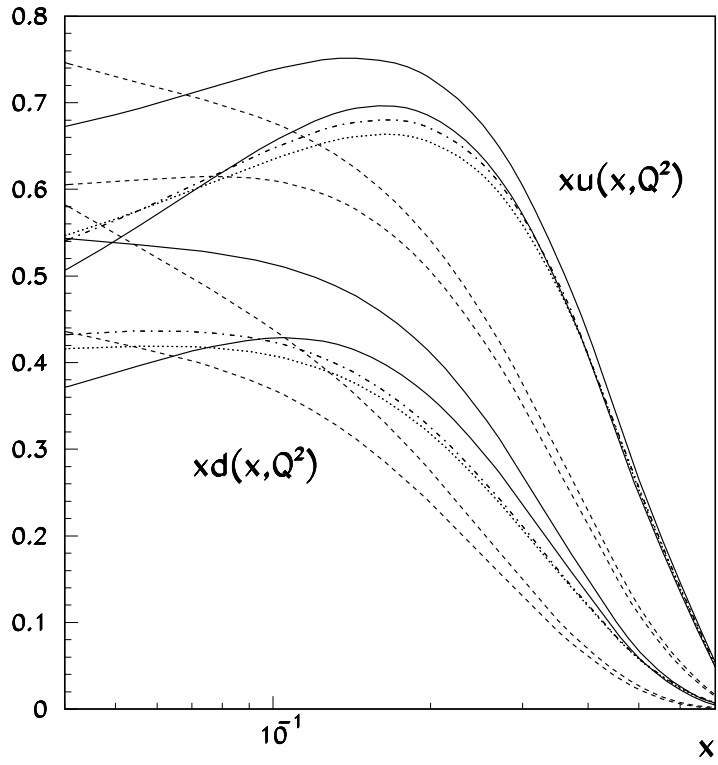 scaled 1000}
\caption{The same as in Fig.6 for the valence quarks.}
\end{center}
\end{figure}
\begin{figure}
\begin{center}
\ForceHeight{0.4\textheight}\BoxedEPSF{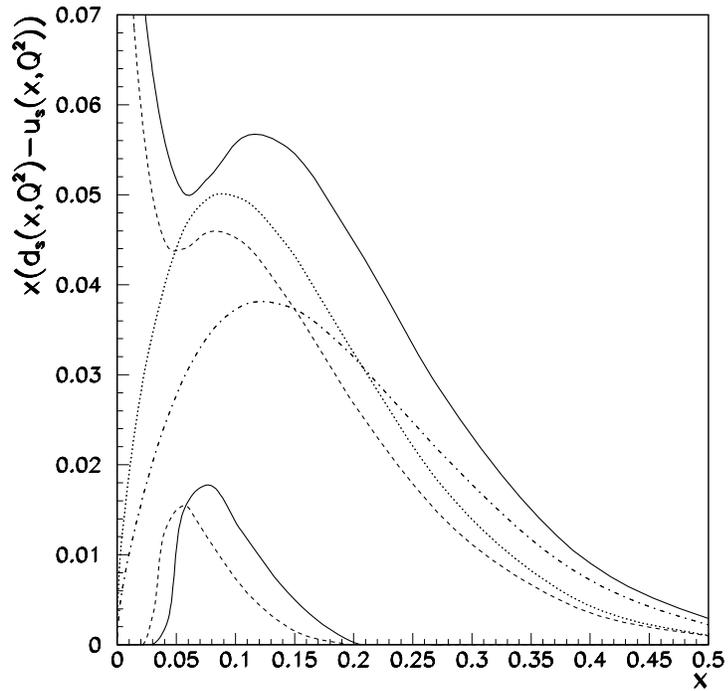 scaled 1000}
\caption{The same as in Fig.6 for the nonstrange sea asymmetry.}
\end{center}
\end{figure}
One could  achieve this due 
\clearpage
to the measurement of $F_2$ at small $x$, which defines gluon
distribution in this region and provides momentum constraint to 
determine it at larger $x$ as well. 
As to the quark distributions they are determined much more precisely.
We should, however, point out that 
the obtained PDFs and their errors are certainly model dependent. 
Say, releasing the condition 
$a_{su}=a_{sd}=a_{ss}$ significantly increases the 
errors of sea distributions at the small $x$. Analogous effect
arises if one adds more polynomial terms to the initial PDFs.
The model dependence is inevitable in such analysis
since one cannot determine the continual
functional form of a distribution 
having the limited set of measurements and without additional constraints.
In our case this model dependence is more pronounced for
the quark distributions because the considered data are well known
to have limited potential in the discrimination of sea and valence quarks
meanwhile the gluon distribution and  $\alpha_s(M_Z)$ are less model dependent. 
It is well understood as far as the latter are defined from  
the $F_2$ derivatives, less sensitive to the variation of 
separate quark distributions. 
At small $x$ and large $Q$ one can observe shrinking the error
bands of gluon distribution. This reflects a well known property 
of the DGLAP equation 
based on the dominance of the singular terms and leading to the focusing 
of any input gluon distribution to the universal form \cite{GRIBOV}.

For the comparison we also present the parametrizations MRS(R1) \cite{MRS}
and CTEQ4M \cite{CTEQ} on these figures.
This comparison is limited because of the lack of the error bands for
their PDFs, but any way is more conclusive than the comparison of two
curves without any error bands moreover that one can suppose
the error bands for MRS and CTEQ PDFs to be smaller 
than ours since these groups use more data in the fit. 
We observe 
the statistically significant difference of our gluon distribution 
with those given by MRS and CTEQ sets at 
large $x$, which can be ascribed to
using in these analysis the data on prompt photon 
production. 
The interpretation of these data has been recently recognized to suffer from 
the large ambiguities \cite{PHOTON} and 
the alternative analysis of prompt photon data with the improved theoretical 
treatment of these ambiguities give much lower gluon distribution at
moderate $x$ \cite{VOGEL}, compatible with ours. 
As to the discrepancies in 
$d$-quark and, to less extent, $u$-quark distributions at moderate $x$,
the additional investigation showed that they are partially explained
by the influence of target mass and Fermi motion corrections.
One can note that
the larger values of quark distributions in this region of $x$
can help to explain the excess of
the recent data on jet production from 
Fermilab collider over NLO QCD predictions 
in the region of $E_T=200-400~GeV$, where the basic contribution
comes from quark-quark scattering (viz \cite{MRS,CTEQ}).
 In addition to the above, all these 
discrepancies can also originate due to the possible
numerical inaccuracies in 
MRS's and CTEQ's QCD evolution codes reported recently \cite{NLOCODES}
and difference in the $\alpha_s$ values.
The difference in the sea value at $x\sim0.3$ seems to be statistically 
insignificant and can disappear
after inclusion of more data in the analysis.

For the value of $\alpha_s(M_Z)$ the robust estimate is
\begin{displaymath}
\alpha_s(M_Z)=0.1146\pm0.0036~(75\%~C.L.),
\end{displaymath}
compatible with \cite {VIR}, but 
less sensitive to the higher twist contribution.
This estimate is not essentially biased
if the PDFs functional form is changed from (\ref{GENPDF})
to our final form and hence we
can conclude that these estimates are, in the good approximation,
model independent.

\section{Conclusion}

 The Bayesian treatment of systematic errors is the clear and 
efficient method in the analysis of data with numerous sources of
systematic errors and in particular data on DIS scattering.
 This approach allows for a straightforward and correct
account of point-to-point correlations contrary to 
widely used `simplification' consisting of combining 
statistic and systematic errors in quadrature.
The certain suspicions that the estimator using covariance matrix
suffer from the bias proved to be irrelevant  
if one uses the estimator inspired by the maximum likelihood function.
First time the quark and gluon distributions from the global fit
with the full account of experimental errors are obtained.
These PDFs can be extremely useful for further 
phenomenological studies.
Having estimation of PDFs' error bands one can conclusively 
compare the results of various global fits, PDFs extracted from 
different processes and evaluate the statistical significance of 
theoretical uncertainties in the fitted formula. At last, 
calculation of cross sections for other processes, based on 
PDFs are more meaningful if one can account for PDFs'
uncertainties. 

\section{Acknowledgements}

The author expresses his acknowledgment to R.M.Barnett and P.S.Gee for 
providing computing facilities, R.Eichler and J.Feltesse for the 
supply of experimental data and valuable comments, 
L.W.Whitlow for providing the code for calculation of $R_{SLAC}^{1990}$
and A.S.Nikolaev for the code to calculate the Paris wave function 
of deuterium.


\begin{thebibliography}{50}

\bibitem{SOPER}
    Soper Davison E., Collins John C. -- CTEQ NOTE 94/01, 1994,
                      [hep-ph/9411214];\\
    Soper Davison E., Talk given at 30-th Moriond, Meribel les Allues,
    France, 18-25 Mar 1995, [hep-ph/9506218].

\bibitem{MRS}
  Martin A.D., Roberts R.G., Stirling W.J., //Phys. Rev. 1994, V.D50, P.6734;\\
  Martin A.D., Roberts R.G., Stirling W.J. -- DTP/96/44, May 1996,
                      [hep-ph/9606345].

\bibitem{GRV}
    Gl\"{u}ck M., Reya E., Vogt A., //Z. Phys. 1995, V.C67, P.433.

\bibitem{CTEQ}
    CTEQ collaboration, Lai H.L. et. al., //Phys. Rev. 1995, V.D51, P.4763;\\
    CTEQ collaboration, Lai H.L. et. al. -- MSUHEP-60426, June 1996,
                      [hep-ph/9606399].
     
\bibitem{MINUIT}
James F., Roos M. -- CERN Program Library D506. MINUIT --
 Function Minimization and Error Analysis, Version 92.1, 1992.

\bibitem{STAB}
  Alekhin S.I. -- IHEP 94-70, Protvino, 1994, [CERN-SCAN-9501137].

\bibitem{DISP}
  Alekhin S.I. -- IHEP 95-65, Protvino, 1995, [CERN-SCAN-9508274];\\
  Alekhin S.I. -- IHEP 95-48, Protvino, 1995, [CERN-SCAN-9511190].

\bibitem{AGO} D'Agostini G., //Nucl. Instr. and Meth. in Phys. Res. 1994,
                      V.A346, P.306.

\bibitem{BCDMS}
BCDMS collaboration, Benvenuti A.C. et al., //Phys. Lett. 1989, V.223B, P.485;\\
BCDMS collaboration, Benvenuti A.C. et al., //Phys. Lett. 1990, V.237B, P.592.

\bibitem{NMC}
 NM collaboration, Arneodo M. et al. -- [hep-ph/9610231], submitted to 
 Nucl. Phys B.

\bibitem{ZEUS}
 ZEUS collaboration, Derrick M. et al. -- DESY 96-076, June 1996,
                      [hep-ex/9607002].

\bibitem{H1}
 H1 collaboration, Aid S. et al. -- //Nucl. Phys. 1996, V.470B, P.3.

\bibitem{CCFR}
CCFR collaboration, Bazarko et al., //Z. Phys. 1995, V.C65, P.189.

\bibitem{AP}
Gribov V.N., Lipatov L.N., //Sov. Journ. Nucl. Phys. 1972, V.15, P.438;
ibid. P.675;\\
Altarelli G., Parisi G., //Nucl. Phys. 1977, V.126B, P.298;\\
Dokshitzer Yu. L., //Sov. Phys. JETP 1977, V.46, P.641. 

\bibitem{MS}
 Furmanski W., Petronzio R., //Z. Phys. 1982, V.C11, P.293;\\
 Curci G., Furmanski W., Petronzio R., //Nucl. Phys. 1980, V.B175, P.27;\\
 Furmanski W., Petronzio R., //Phys. Lett. 1980, V.97B, P.437.
 
\bibitem{GLUCK}
 Gl\"{u}ck M., Reya E., Stratmann M., //Nucl. Phys. 1994, V.B422, P.37.

\bibitem{NLOCODES}
 Bl\"{u}mlein J., Riemersma S., Botje M. et al., -- DESY 96-199, Sep. 1996,
                      [hep-ph/9609400].

\bibitem{PDG}
 Particle Data Group, //Phys. Rev. 1996, V.54, P.77.

\bibitem{TMC}
 Georgi H., Politzer H.D., //Phys. Rev. 1976, V.D14, P.1829.

\bibitem{LOWQ}
    Martin A.D., Roberts R.G., Stirling W.J., //Phys. Rev. 1995, V.D51, P.4756.

\bibitem{WHIMP}
 Bazizi K., Whimpenny S.J. -- UCR/DIS-90-04, 1990.

\bibitem{WHITLOW}
Whitlow L.W. et al., // Phys. Lett. 1990, V.250B, P.193.

\bibitem{WEST}
West Geoffrey B., // Ann. Phys. 1972, V.74, P.464. 

\bibitem{PARIS}
Lacombe M. et al., // Phys. Rev. 1980, V.C21, 861;\\
Lacombe M. et al., // Phys. Lett. 1981, V.101B, 139.

\bibitem{SOKOL}
Sokolov S.N. -- IFVE 88-110, Protvino, 1988.

\bibitem{NMCGLU}
 NM collaboration, Arneodo M. et al., // Phys. Lett. 1993, V.309B, P.222.

\bibitem{GRIBOV}
Gribov V.L., Levin E.M., Ryskin M.G., // Phys. Rep. 1981, V.C100, P.1. 

\bibitem{PHOTON}
    CTEQ collaboration, Huston J. et. al., // Phys. Rev. Lett. 1996, 
                                                   V.77, P.444.
\bibitem{VOGEL}
    Vogelsang W., Vogt A., // Nucl. Phys. 1995, V.B453, P.334.

\bibitem{VIR}
   Virchaux M., Milsztajn A., //Phys. Lett. 1992, V.274B, P.221.

\end{thebibliography}
\end{document}